# Tuning Conductivity Type in Monolayer $WS_2$ and $MoS_2$ by Sulfur Vacancies

Jing Yang[1*], Fabio Bussolotti[2], Hiroyo Kawai[1] and Kuan Eng Johnson Goh[2,3**]

[1] *Institute of High Performance Computing, A*STAR (Agency for Science, Technology and Research), 1 Fusionopolis Way, #16-16 Connexis, Singapore 138632*
[2] *Institute of Materials Research and Engineering, A*STAR (Agency for Science, Technology and Research), 2 Fusionopolis Way, Innovis, #08-03, 138634 Singapore*
[3] *Department of Physics, Faculty of Science, National University of Singapore, 2 Science Drive 3, 117551 Singapore*

*Email: yang_jing@ihpc.a-star.edu.sg
**Email: kejgoh@yahoo.com;   gohj@imre.a-star.edu.sg

**ABSTRACT**

While *n*-type semiconductor behavior appears to be more common in as-prepared two-dimensional (2D) transition metal dichalcogenides (TMDCs), substitutional doping with acceptor atoms is typically required to tune the conductivity to *p*-type in order to facilitate their potential application in different devices. Here, we report a systematic study on the equivalent electrical "doping" effect of - single sulfur vacancies ($V_{1S}$) in monolayer $WS_2$ and $MoS_2$ by studying the interface interaction of $WS_2$-Au and $MoS_2$-Au contacts. Based on our first principles calculations, we found that the $V_{1S}$ can significantly alter the semiconductor behavior of both monolayer $WS_2$ and $MoS_2$ so that they can exhibit the character of electron acceptor (*p*-type) as well as electron donor (*n*-type) when they are contacted with gold. For relatively low $V_{1S}$ densities (approximately < 7% for $MoS_2$ and < 3% for $WS_2$), the monolayer TMDC serves as electron acceptor. As the $V_{1S}$ density increases beyond the threshold densities, the $MoS_2$ and $WS_2$ play the role of electron donor. The significant impact

$V_{1S}$ can have on monolayer $WS_2$ and $MoS_2$ may be useful for engineering its electrical behavior and offers an alternative way to tune the semiconductor TMDCs to exhibit either *n*-type or *p*-type behavior.

## 1. INTRODUCTION

As a member of transition metal dichalcogenides (TMDCs), two-dimensional (2D) tungsten disulfide ($WS_2$) has attracted enormous interests due to its attractive properties for potential applications in electrical and optical field,[1] especially in the field of low power FETs,[2] photodectors,[3] gas and chemical sensors,[4] memory and electroluminescent devices, and integrated circuits.[5] Pristine $WS_2$ is reported to be an *n*-type semiconductor. Ovchinnikov *et. al* fabricated FETs based on single layers semiconductor $WS_2$ and reported *n*-type behavior with a high on/off current ratio of ~$10^6$ at room temperature.[6] In addition to *n*-type behavior, Morrish *et. al* reported that semiconducting $WS_2$ layers showed an indirect band gap of 1.4 eV and an absorption coefficient of ~ $5 \times 10^4$ $cm^{-1}$.[7]

To realize the attractive properties in potential applications, significant efforts have been directed to tuning the properties of $WS_2$. Doping, which can controllably tailor both the material's crystalline structure as well as band gap, is one of the most common procedures that has been widely demonstrated both theoretically and experimentally. Iqbal *et. al* showed an unprecedented high mobility of 255 $cm^2$ $V^{-1}$ $s^{-1}$ at room temperature in a $WS_2$ FET fabrication by *n*-type chemical doping of $WS_2$ films.[8] Shi *et. al* synthesized Co-doped $WS_2$, and they used both calculations and experiments to demonstrate that the Co-doped $WS_2$ was more active for the hydrogen evolution reaction compared to pristine $WS_2$.[9] Azizi *et. al* synthesized striped triangular monolayer flakes of $W_xMo_{1-x}S_2$ and found in them both spin-orbit and thermal transport anisotropies.[10]

For semiconducting TMDCs, many recent works have reported effective ways for *n*-type doping, such as hydrogen adsorption on TMDCs[11] and chemisorption of oxygen on surface defect $MoS_2$[12] induces the n-doping effect. It is also important to effectively dope the TMDCs as *p*-type for their potential application in 2D-semiconductor-based CMOS-FETs. Currently, most TMDC logic circuits are based only on *n*-channel metal–oxide–semiconductor FETs.[13] The *p*-type TMDC is also important in photodiodes and light-emitting devices as well as other *p-n* junction-based optoelectronic devices.[14-16] Jin *et. al* demonstrated that substitutional doping of $WS_2$ with Nb changes the carrier type, switching the $WS_2$ transport behavior from intrinsic *n*-type to *p*-type. In addition, it was found that the bandgap and their optical properties can be tuned by carefully controlling the Nb density.[17] Laskar *et. al* showed that Nb can also be used as *p*-type dopant in $MoS_2$.[18] Tang *et. al* reported that the electrical characteristics of monolayer/few-layer $WS_2$ FETs clearly show an *n*-channel to *p*-channel conversion with nitrogen incorporation.[19] Similarly, Azcatl *et. al* demonstrated that *p*-type doping of $MoS_2$ is attained by substitutional nitrogen doping.[20] Furthermore, various reports have shown that *p*-type doping behavior can also be obtained by introducing carbon to monolayers of $WS_2$[21], phosphorus to $MoS_2$[22] and oxygen adsorption to $MoS_2$[23]. However, these doping methods pose major concerns in terms of crystal stability, impurity, and structural damage (or even etching) by energetic ions, particularly for monolayers.[24-25] Hence, recent works have also investigated alternative methods for tuning the electrical transport properties of TMDC monolayers without the introduction of foreign atomic species. One example is to employ the charge transfer arising from different type/density of sulfur vacancies on monolayer $MoS_2$ to create a *p*-type or *n*-type semiconductor/metal contact[26].

We present herein a method to endow monolayer $WS_2$ or $MoS_2$ with both electron acceptor (*p*-type) and electron donor (*n*-type) behavior in conjunction with the use of Au as contact to the monolayer TMDC. A single sulfur vacancy ($V_{1S}$) is introduced into unit cells of different sizes to mimic different $V_{1S}$ defect densities. We find that even though the intrinsic $WS_2$ behaves like an *n*-type semiconductor once in contact with metal, the introduction of 1.6% $V_{1S}$ (density of $1.8 \times 10^{13}$ cm$^{-2}$) can induce a charge transfer behavior similar to a *p*-type doping. As the density of $V_{1S}$ is increased to $7.1 \times 10^{13}$ cm$^{-2}$ or 6.3% and beyond, it switches again to an *n*-type behavior. The charge transfers associated with such $V_{1S}$ appear correlated to the energy offset between the conduction band minimum (CBM) and unoccupied defect state. We find this association to be consistent and generalizable to the case of $MoS_2$ as well. Our results indicate defect engineering as an effective method for controlling the electronic properties of TMDC monolayer interfaced with metals, where the largest area interaction between TMDC MLs and electrodes exist in actual devices, without the need of foreign atom introduction.

## 2. MODELS AND METHODS

All the studies were performed by using the plane-wave technique implemented in Vienna ab initio simulation package.[27-28] The generalized gradient approximation with the Perdew-Burke-Ernzerhof (PBE) functional has been employed to describe the exchange-correlation potential in all calculations.[29-30] While hybrid functionals (e.g. Heyd-Scuseria-Ernzerhof (HSE), PBE0) may be used for obtaining more accurate band structures, they are computationally more demanding than the traditional PBE functional. Considering that the hybrid functionals, such as HSE, PBE0, do not change the general trend,[31-32] the PBE functional was applied in this study as a compromise between the computational cost and accuracy. In order to study how the interface interaction at Au(1 1 1)-WS$_2$ and Au(1 1 1)- MoS$_2$ contacts is influenced by V$_{1S}$ (as shown in Figure 1), 2×2, 4×4, 6×6 and 8×8 supercells were constructed, which mimics the V$_{1S}$ density of 25%, 6.3%, 2.8% and 1.6 %, respectively. Since the convergence test for K-points showed that the charge transfer result is not sensitive to different K-points, 2×2×1 K-point mesh was used for all Au-WS$_2$ and Au-MoS$_2$ systems. For isolated WS$_2$ and MoS$_2$, we applied a 9×9×1 K-point mesh for more accurate band structure analysis. The cutoff energy was set to be 500 eV. A 25 Å vacuum layer was used to separate the periodic slab in vertical direction for all interface structures. The van der Waals interactions is corrected with Grimme's dispersion-corrected density functional theory (DFT-D2) method.[33] All structures were optimized until the Hellmann−Feynman force on each atom was smaller than 0.05 eV/Å.

We used electron density difference map and Bader charge transfer amount to evaluate the contact-TMDC interaction at the interface. The electron density difference map is defined as the electron density difference before and after the contact between Au and TMDCs:

$$\Delta\rho = \rho_{Au-MoS2} - \rho_{Au} - \rho_{MoS2}$$

so that it can directly visualize the direction and amount of electron movement. The Bader charge calculations were carried out to evaluate the amount of charge transferred between Au and TMDC. If electrons are transferred from the metal to the TMDC upon contact, the TMDC acts as an electron acceptor, which corresponds to *p*-type behavior. On the other hand, if electrons are transferred from the TMDC to the metal, the TMDC acts as an electron donor, which corresponds to *n*-type behavior.

To gain more insight into the effect of $V_{1S}$ we also determined the energy offset ΔE defined as the energy difference between the CBM and the maximum of the unoccupied defect state:

$$\Delta E = E_{CBM} - E_{\text{unoccupied defect state maximum}}$$

## 3. Results and discussion

The Au-WS$_2$ and Au-MoS$_2$ interfaces were firstly evaluated by physical separation and then by the electron density difference and Bader charge calculations. These are followed by a detailed study on the band structure of monolayer WS$_2$ and MoS$_2$, focusing on the effect of various $V_{1S}$ densities.

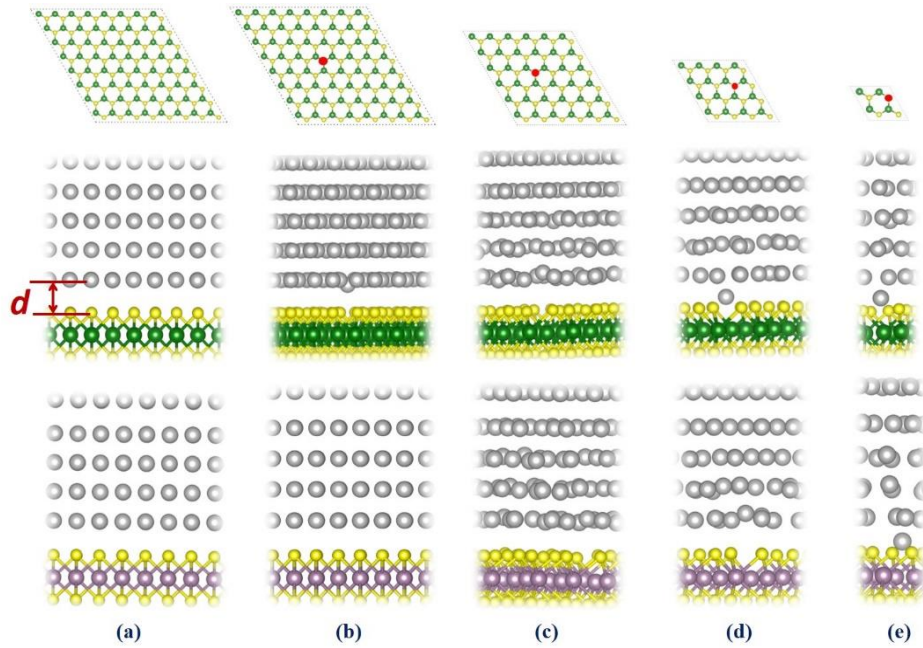

**Figure 1.** The optimized structures of Au-WS$_2$ and Au-MoS$_2$ contacts with various V$_{1S}$ densities (a) 0%, (b) 1.6%, (c) 2.8%, (d) 6.3% and (e) 25.0%. The top panel represents supercell structures and middle panel represents the Au-WS$_2$ interfaces and bottom panel represents the Au-MoS$_2$ interface. Yellow, grey, purple and green spheres represent S, Au, Mo and W atoms, respectively. As shown in (a), the distance between the topmost S layer and the bottom Au layer is defined as the interface thickness.

The optimized structures of the Au-WS$_2$ and Au-MoS$_2$ interface systems with various V$_{1S}$ densities are shown in Figure 1. The physical separation $d$, as shown in Figure 1(a), is defined as the thickness of the interface. A thinner interface with smaller $d$ value suggests a potentially more orbital overlap at the interface and hence the formation of stronger bond, such as covalent bond. For Au and pristine WS$_2$, the calculated $d$ is 2.37 Å which is 0.11 Å larger than the sum of the S and Au covalent radii, resulting in a limited orbital overlap and very weak interaction between WS$_2$ and Au. This is in line with a previous study which claims that the interaction between Au and WS$_2$ is considered as physical adsorption and there is no obvious electron overlap between the bottom S atoms and top Au atoms.[34] Similar weak interaction is also detected in the Au-MoS$_2$ contact.[35-36] Upon the creation of the V$_{1S}$, there are rearrangements of the Au atoms as shown in Figure 1(b) to (e). The rearrangement of interface Au atoms becomes more significant as the V$_{1S}$ density increases from 1.6% to 25.0% for both Au-WS$_2$ and Au-MoS$_2$.

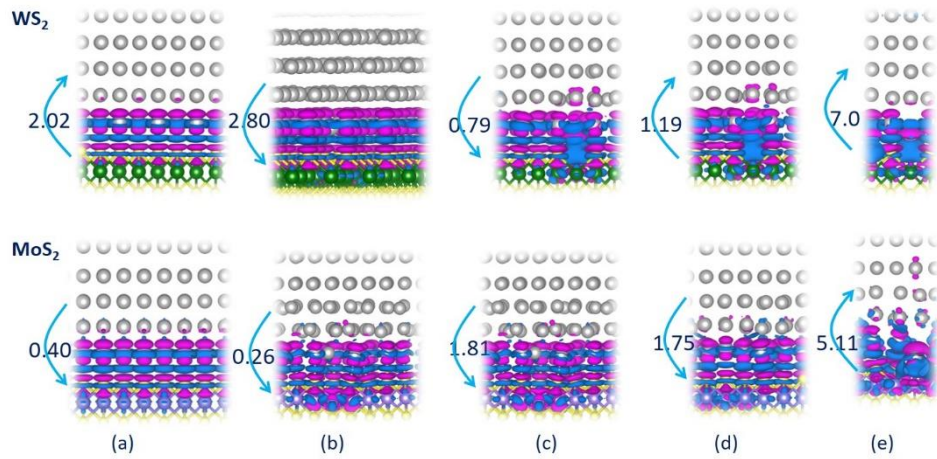

**Figure 2. Electron density difference map at Au-WS$_2$ and Au-MoS$_2$ interfaces with various V$_{1S}$ densities (a) 0%, (b) 1.6%, (c) 2.8%, (d) 6.3% and (e) 25.0%. The top panel represents the Au-WS$_2$ interface and bottom panel represent for Au-MoS$_2$ interface. Yellow, grey, purple and green spheres represent S, Au, Mo and W atoms, respectively. The blue and purple zones correspond to electron accumulation and depletion areas, respectively. Each blue arrow and the corresponding number indicate the charge transfer direction and amount ($10^{-2}$ $e^-$) predicted by Bader charge, respectively. (Isosurface = 0.002 electron·Å$^{-3}$)**

We carried out the electron density difference map and charge transfer amount calculations to evaluate the Au-WS$_2$ and Au-MoS$_2$ contact interfaces. Figure 2(a) top panel shows the electron density difference map associated with the interaction between pristine WS$_2$ and Au. Although there is electron movement upon interaction, it is not sufficient to conclude the overall electron transfer direction. In order to investigate on the charge transfer in detail, Bader charge calculation was performed. The results suggest that the electron transfer direction is from WS$_2$ to Au and the charge transfer amount is $2.0\times10^{-2}$ $e^-$ per surface Au atom. In other words, the monolayer WS$_2$ serves as electron donor leading to an *n*-type contact with Au. This is in line with the experimental FET results which identified electrons as majority carriers in pristine monolayer WS$_2$.[6-7] The Figure 2(b) top panel shows the electron density difference map associated with the contact between Au and WS$_2$-V$_{1S}$. Upon the creation of V$_{1S}$, charge disorder is created at the interface and the electrons tend to accumulate around the missing S atoms site. This is observed in Figure 2(b) top panel where the charge accumulation and depletion areas (blue and purple, respectively) are slightly distorted compared to the pristine case and some charge accumulations are seen in WS$_2$ layer. The electron movements from Au to WS$_2$-V$_{1S}$ indicate that the WS$_2$ with a density of 1.6% V$_{1S}$ serves as an electron acceptor and thus the V$_{1S}$ induces an overall *p*-type behavior in the monolayer WS$_2$. The amount of charge transfer evaluated by the Bader charge analysis is $2.8\times10^{-2}e^-$ per surface Au atom. This introduction of a relatively small number of defects can tune its conductivity type to electron acceptor with respect to the defect-free WS$_2$. A possible reason for its *p*-type behavior is that the deep defect states created by the sulfur vacancies may act as electron-trap centers.[37-39] A similar effect is detected when V$_{1S}$ density is further increased to 2.8%, as shown in Figure 2(c) top panel. However, the charge transfer amount is reduced to $0.79\times10^{-2}e^-$ per surface Au atom, indicating a reduced *p*-type behavior compared to 1.6% V$_{1S}$. When the V$_{1S}$ density is increased to 6.3%, as shown in Figure 2(d) top panel, the electron movement becomes more substantial but in the opposite direction. Instead of accumulating around the defect site, the electron tends to migrate towards Au layers, indicating that the $WS_2$-$V_{1S}$ act as electron donor and the $V_{1S}$ appears to have an effect similar to an *n*-type doping. The amount of charge transfer is calculated to be $1.19\times10^{-2} e^-$ per surface Au atom. One possible explanation is that as the unsaturated electron associated with the sulfur vacancy is located on neighboring W atoms, the number of unsaturated electrons on W atoms increase when the defect density is increased. This makes the $WS_2$ electron rich and serve as the source of electron carriers. This explanation is also applicable for Au-$WS_2$ with 25% $V_{1S}$. With more missing S atoms, there are more unsaturated electrons and make the $WS_2$ more electron-rich. As shown in Figure 2(e), the relevant amount of charge transferred from $WS_2$ to Au increases significantly to $7.0\times10^{-2} e^-$ per surface Au atom. This is in line with previous work which showed that the sulfur vacancy produces an *n*-type doping effect.[40]

We have performed a similar analysis for the effect of $V_{1S}$ on $MoS_2$. As shown in Figure 2(a) bottom panel, there is hardly any charge redistribution upon the contact of Au and pristine $MoS_2$, indicating a weak interaction between them. $V_{1S}$ makes the $MoS_2$ electron-poor when its density is 1.6%, and this trend continues to a density of 2.8 %. Although the charge transfer amount slightly increases for $V_{1S}$ density of 6.8 %, the electron transfer direction is still from Au to $MoS_2$. This is in line with the report that the native sulfur vacancies in $MoS_2$ generate deep levels below the CB minimum and make it act as an electron-trap center.[38] When the $V_{1S}$ density is increased to 25%, the $MoS_2$ becomes electron donor, suggesting that the $V_{1S}$ is producing an effectively *n*-type doping behavior. This agrees with previous work claiming that the sulfur vacancy is a sufficiently shallow electron donor state resulting in *n*-type monolayer $MoS_2$.[41] In addition, theoretical evidence supports that sulfur vacancies that exist in $MoS_2$ introduce localized donor states inside the bandgap.[42] The fact that $MoS_2$ can switch its role between electron acceptor and electron donor agrees with our previous study[26] as well as the experimental study reporting that $MoS_x$ experienced a transition from *p*-type to *n*-type as the ratio of S:Mo decreases.[43]

From the electron density difference map and Bader charge calculation, we therefore find that the charge transfer direction between $WS_2$ or $MoS_2$ and Au can be tuned by varying the $V_{1S}$ density. In other words, the "doping" effect can be switched between *p*-type and *n*-type behaviors for both $WS_2$ and $MoS_2$ by tuning their sulfur vacancy density. The amount of charge transfer per surface Au atom as a function of the $V_{1S}$ density is summarized in Figure 3. Once in contact with Au, the pristine $WS_2$ tends to acquire charge from the metal layers, thus behaving as an *n*-type semiconductor, while the pristine $MoS_2$ is an intrinsic semiconductor. Upon the creation of a relatively small amount of $V_{1S}$, both $WS_2$ and $MoS_2$ become electron acceptor, inducing a *p*-type semiconductor behavior. As the $V_{1S}$ density is increased beyond their respective threshold densities, both exhibit *n*-type behavior. In general, for a highly defective system, $V_{1S}$ leads to the $WS_2$ or $MoS_2$ having an electron donor (*n*-type) behavior, while for a system with sufficiently low defect densities, $V_{1S}$ can induce a *p*-type behavior. The crossover $V_{1S}$ density is dependent on the TMDC material.

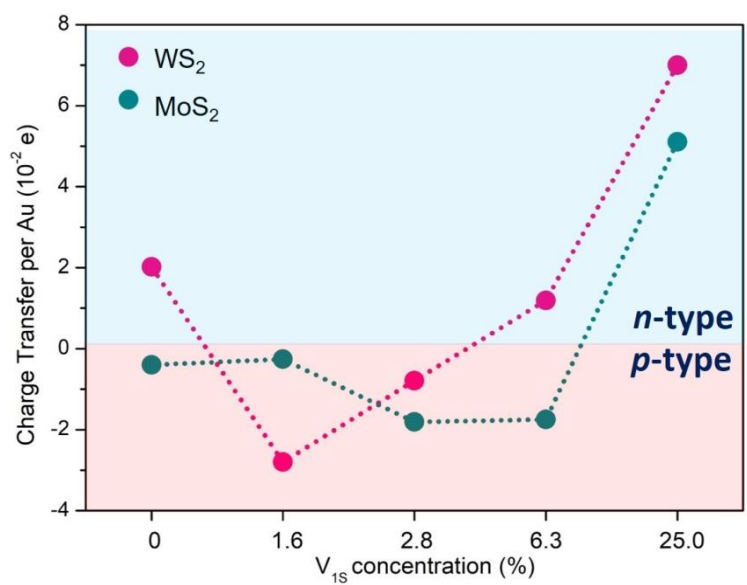

**Figure 3. The effect of V$_{1S}$ density in MoS$_2$ and WS$_2$ on interface charge transfer.** The charge transfer amount is represented as per surface Au atom to allow comparison between different defect densities. Green and pink dots correspond to Au-MoS$_2$ and Au-WS$_2$ interfaces, respectively. A positive charge transfer amount indicating the electron movement from TMDCs to the metal and suggesting an *n*-type behavior of the TMDCs, and a negative charge transfer denotes the opposite direction, indicating a *p*-type behavior of the TMDCs.

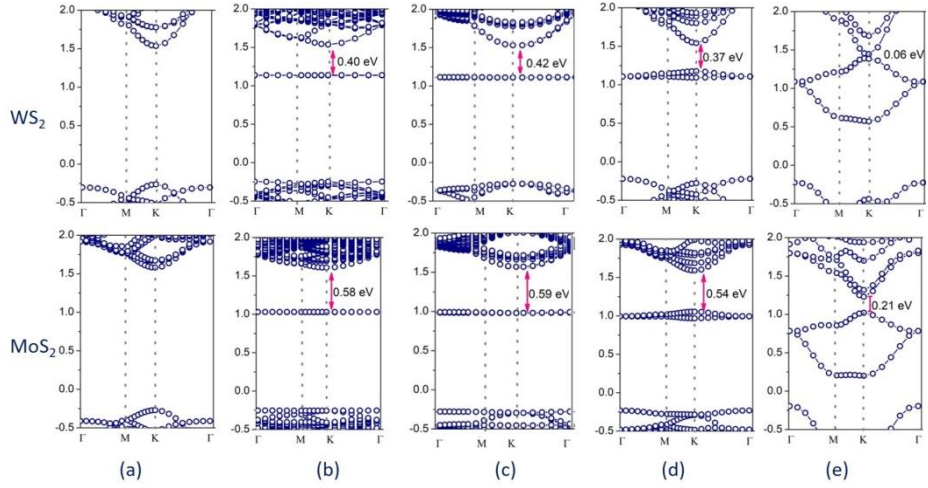

**Figure 4.** Band structure of WS$_2$ (top panel) and MoS$_2$ (bottom panel) with various V$_{1S}$ densities (a) 0%, (b) 1.6%, (c) 2.8%, (c) 6.3% and (d) 25.0%. The red arrow and the number (in eV) show the energy level offset between unoccupied defect state and CB bottom for each band structure.

To further understand why different V$_{1S}$ densities can produce *p*- or *n*-type behavior, a detailed study on the band structure was performed, focusing on the defect states. The band structures of monolayer WS$_2$ and MoS$_2$ with various V$_{1S}$ densities are summarized in Figure 4. The band gap of pristine monolayer WS$_2$ is direct and about 1.80 eV, which agrees well with the literature.[21] The existence of V$_{1S}$ results in the generation of defect states at the band gap regions (as shown in Figure 4, b-e). With a V$_{1S}$ density of 1.6%, an unoccupied defect state is introduced, with an energy offset of 0.40 eV between the CBM and the top of unoccupied defect state. The deep defect states created by the sulfur vacancies prohibit them from acting as electron donors so that they do not contribute to more *n*-type behavior.[37] As the V$_{1S}$ densities increases, these two unoccupied defect states gradually split into two and the energy offset changes to 0.42 eV (2.8%), 0.37 eV (6.3%) and 0.06 eV (25.0%) for each V$_{1S}$ density. For pristine monolayer MoS$_2$, it has a direct band gap of about 1.84 eV, which is in line with previous DFT calculations.[38] At the same $V_{1S}$ density, the absolute energy offset of the defect state is larger for MoS$_2$ than that of WS$_2$ while the trend is the same. The energy offset is 0.58 eV for a $V_{1S}$ density of 1.6%. As the density of $V_{1S}$ is increased, the unoccupied defect states split, and the energy offset becomes 0.54 eV (2.8%), 0.59 eV (6.3%) and then reduces to 0.21 eV (25.0%).

The amount of charge transfer (per surface Au atom) as a function of unoccupied defect state energy offset ( $\Delta E = E_{CBM} - E_{\text{unoccupied defect state maximum}}$ ) is summarized in Figure 5. For WS$_2$ with a $V_{1S}$ density of 1.6% and 2.8%, large energy offsets of 0.40 and 0.42 eV are observed respectively. These deep unoccupied defect states below the CBM might prohibit them from acting as electron donor but makes them more conducive and act as electron acceptor (*p*-type behavior). For higher $V_{1S}$ densities (6.3% and 25.0%), smaller energy offset is observed, which favors an *n*-type behavior. A similar trend is found in MoS$_2$: for *n*-type behavior (density of 25% $V_{1S}$), the energy offset is small (0.21 eV); for *p*-type behavior (density 1.6%, 2.8% and 6.3% $V_{1S}$), the energy offsets are relatively large (0.58eV, 0.59eV and 0.54 eV). A similar reasoning may explain why the 6.3% density $V_{1S}$ case demonstrates an *n*-type behavior in WS$_2$ but *p*-type in MoS$_2$. In MoS$_2$, the unoccupied defect state of 6.3% $V_{1S}$ is deep below the CBM, about 0.54 eV from the CBM. The unoccupied state is sufficiently low in energy so that it favors electron injection. One the contrary, the unoccupied defect state of 6.3% $V_{1S}$ is shallower in WS$_2$ (0.37 eV below the CBM) which might explain its *n*-type behavior.

While controlling the doping level, the increase of defect density can also impact on the charge mobility, as sulphur vacancies may act as scattering center for the carriers injected into MLs. Therefore a balance between these two effects must be carefully considered in designing TMDC-based device.

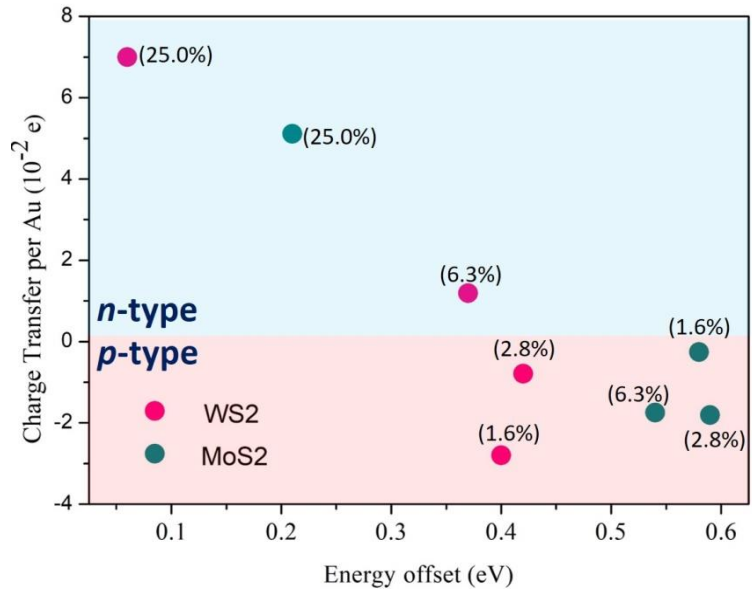

**Figure 5. The amount of charge transfer (per surface Au atom) as a function of the energy offset between the CB bottom and the unoccupied defect states. Green and pink dots correspond to Au-MoS$_2$ and Au-WS$_2$ contacts, respectively.**

## 4. CONCLUSION

We have performed a systematic study on the equivalent electrical "doping" effect of single sulfur vacancies in monolayer WS$_2$ and MoS$_2$ by studying the interfacial interactions of WS$_2$-Au and MoS$_2$-Au contacts. Different defect densities were explored. Both monolayer MoS$_2$ and WS$_2$ demonstrate the possibility to switch between electron acceptor (*p*-type) and electron donor (*n*- type) behavior when the V$_{1S}$ density is varied. For relatively low V$_{1S}$ densities (approximately < 7% or 7.2 x 10$^{13}$ cm$^{-2}$ for MoS$_2$, and < 3% or 3.1 x 10$^{13}$ cm$^{-2}$ for WS$_2$), the monolayer TMDC serves as electron acceptor, suggesting that the $V_{1S}$ imparts to the TMDC a *p*-type behavior. As the number of $V_{1S}$ increases beyond the respective threshold density, the $MoS_2$ and $WS_2$ play the role of electron donor which suggests that the $V_{1S}$ imparts *n*-type behavior to the TMDC. We use the energy offset ($\Delta E$) from the TDMCs CBM to the unoccupied $V_{1S}$ defect state to obtain more insight on the effect of varying the density of $V_{1S}$. When the $\Delta E$ is smaller, the unoccupied $V_{1S}$ defect state is closer to the TDMCs CBM, and the TMDC tend to act as electron donor (*n*-type behavior). On the other hand, when the $\Delta E$ is larger, the unoccupied $V_{1S}$ defect state is further away from the TDMC CBM which would facilitate electron injection into the TMDC and results in the TMDC acting as the electron acceptor (*p*-type behavior). While we anticipate that real systems will be more complex, the possibility to induce both *p*- and *n*-type electrical behavior via $V_{1S}$ in monolayer $MoS_2$ and $WS_2$ offers a potential pathway to tune conductivity without the introduction of foreign atoms and this work provides first-order guidance as to how one might engineer these behaviors in TMDC materials.

**ACKNOWLEDGMENTS**

This work is supported by the Agency for Science, Technology and Research (A*STAR) under its A*STAR Pharos Grant Nos. 1527000016 and 1527000017. The author acknowledges the computing resources support from A*STAR Computational Resource Centre and National Supercomputer Centre, Singapore.